\documentclass[]{aa}
\input psfig.tex

\newcommand{\ms}{M$_{\odot}$}
\newcommand{\zs}{Z$_{\odot}$}

\newcommand{\N}{N$_{\rm HI}$}
\newcommand{\Nz}{N$_{\rm Zn}$}
\begin{document}

\thesaurus{ }

\title{Chemical evolution and depletion pattern in Damped Lyman $\alpha$ systems }

\author{J.L. Hou \inst{1,2,3,4},
     S. Boissier \inst{2} and 
     N. Prantzos \inst{2} }

\institute{
Shanghai Astronomical Observatory, CAS, Shanghai, 200030, P.R. China
(hjlyx@center.shao.ac.cn) 
\and Institut d'Astrophysique de Paris, 98bis, Bd. Arago, 75014, Paris, France
(prantzos@iap.fr)
\and National Astronomical Observatories, CAS, P.R. China 
\and Joint Lab of Optical Astronomy, CAS, P.R. China 
}
\date {}
\maketitle 


\begin{abstract}

Dust depletion plays a key role in understanding the nature of 
Damped Lyman $\alpha$ systems(DLAs). 
In this paper we point out a previously unnoticed  anticorrelation 
between the observed abundance ratio [X/Zn] (where Zn is assumed to be 
undepleted and X stands for the refractories Fe, Cr and Ni)  and 
metal column density ([Zn/H]+log(N$_{HI}$)) in DLAs. 
We suggest  that this trend is an unambiguous sign of dust depletion, 
since metal column density is a measure of the amount of dust 
along the line of sight. 
Assuming that DLAs are (proto-)galactic disks and using detailed
chemical evolution models with metallicity dependent yields we study 
chemical evolution  and dust depletion patterns for $\alpha$ and iron$-$peak 
elements in DLAs.  When observational constraints on the metal column density
of DLAs are taken into account (as suggested in Boiss\'e et al. 1998) we find
that our models reproduce fairly well the observed mild redshift evolution
of the abundances of 8 elements (Al, Si, S, Cr, Mn, Fe, Zn and Ni) 
as well as the observed scatter at a given redshift. 
By considering the aforementioned dependence of abundance ratios 
on metal column density, we further explore the general dust depletion 
pattern in DLAs, comparing  to our  model results and to a solar reference 
pattern. We find that for low metal column densities (no depletion), 
our models compare fairly well to the data, while a solar pattern has 
difficulties with Mn. At high metal column densities 
(amount of depletion $\sim$0.5 dex), 
the solar pattern describes the data quite well, while our models have 
difficulties with S. We suggest  that further measurements of those key 
elements, i.e. Zn, S and Mn, will help to gain more insight into the 
nature of DLAs. In any case,
the presently uncertain nucleosynthesis of Zn in massive stars
(on which a large part of  these conclusions is based) 
should be carefully scrutinised. 


\end{abstract}

\section{Introduction}

Damped Lyman $\alpha$ systems (DLAs) are high column density gaseous systems 
(\N$>$2 10$^{20}$ cm$^{-2}$), detected through their absorption lines
in the optical spectra of quasars, up to relatively high redshifts
(up to $z\sim$5). Their study constitutes a powerful means to
investigate the properties of distant galaxies (or of their building
blocks). In particular, DLA metal abundances have been widely used
in the past few years to probe the nature of DLAs (e.g.  Lauroesch et al. 1996;
Lu et al. 1996; Pettini et al. 1994, 1997b, 1999, 2000; 
Prochaska \& Wolfe 1999,2000, and many others)).
However, it is not clear whether the observed abundances allow that,
because of various biases: depletion of metals into dust (Pei and Fall 1995)
or bias against too high or too low metal column densities (Boiss\'e 
et al. 1998). The latter bias, in particular, may explain the observed 
absence of evolution in the absolute abundance of Zn/H as a function
of redshift, as suggested in Prantzos and Boissier (2000a, hereafter PB2000a).

Abundance ratios offer a better diagnostic tool than absolute 
abundances for the study of the chemical evolution of a system.
However, in the case of DLAs the possibility of depletion into dust
grains (even to small extent) makes difficult a direct interpretation
of the observed abundance patterns. Several attempts have been made
in recent years to recover the intrinsic abundance pattern from the
observed one and to compare the result to well known patterns of stars
in the Milky Way (e.g. Vladilo 1998, 1999; Pettini et al. 2000), but with rather
modest success up to now.

In this work we reassess the question of abundance patterns in DLAs. First, we 
point out an anticorrelation between
the observed abundance ratio X/Zn (where Zn is assumed to be undepleted
and X stands for the refractories  Fe, Cr and Ni) and metal
column density (Sec. 2). We suggest that this, previously unnoticed, trend is an
unambiguous sign of depletion, since metal column density is
a measure of the amount of dust along  the line of sight. We then study 
the chemical evolution of those systems, assuming they are (proto-)
galactic disks. This assumption is rather controversial, since other
systems (e.g. Low Surface Brightness galaxies, 
dwarf irregulars, galactic haloes;
see Jimenez et al. 1998, Matteucci et al. 1997, Valageas et al. 1999, 
Haehnelt et al. 1998, Petitjean and Ledoux 1999) 
have also been suggested; however, the
various zones of our multi-zone disk models have sufficiently diverse chemical
histories to account for systems other than the Milky Way, a ``prototype''
disk galaxy. Our models, briefly presented in Sec. 3.1, have been shown to
reproduce  successfully  a large number of observed properties of spirals 
at high and low redshifts. An important ingredient of our study is
the use of metallicity dependent yields for massive 
stars and SNIa (presented in Sec. 3.2). In Sec. 4 we show that our results
reproduce fairly well both the observed slow evolution with redshift of the
absolute abundances of several elements, and the corresponding scatter;
observational biases, as already suggested in PB2000a for the case of Zn,
is a key to that success.
In Sec. 5 we compare our results to observed abundance ratios in DLAs, taking
into account the aforementioned trend of depletion vs. metal column density.
We find that for low metal column densities (i.e. no depletion) our models
compare fairly well to the data, while a solar pattern has difficulties with
Mn. At high column densities (small depletion) the solar pattern describes
the data quite well, while our models have difficulties with S. We suggest 
that further measurements of those key elements, i.e. Zn, S and Mn, will
help to gain some insight into the nature of DLAs. In any case,
the presently uncertain nucleosynthesis of Zn (on which
a large part of these conclusions is based) should be carefully scrutinised.

\section {DLA abundance ratios vs. metal column density}

Studies of the elemental abundances of the interstellar medium (ISM) in 
the Milky Way have shown that the gas phase abundances of most elements 
are below the corresponding solar values. This underabundance is generally 
attributed to depletion into dust grains. The amount of depletion depends 
on the nature of the interstellar medium. In cold, dense, disk gas, Mn, 
Cr, Fe and Ni are considerably depleted (by 1.5 to 2 dex), while 
S/H is nearly solar (Savage and Sembach 1996 and references therein). 
In warm disk gas, the amount of depletion is smaller ($\sim$ 1 dex for 
Fe$-$peak elements), while in warm, halo gas, a negligible amount of 
depletion is found ($<$0.5 dex) for all elements, except Fe and Ni which
are depleted by 0.6 to 0.8 dex.

A more quantitative assessement of the amount of depletion can be made by plotting
the abundance of each ionic species as a function of the hydrogen column density
\N \ along the line of sight. In a recent work, 
Wakker and Mathis (2000) compiled data from 
several studies concerning high column densities (\N \ = 10$^{20}$ to 7 10$^{21}$
cm$^{-2}$) and extended that range down to 10$^{18}$ cm$^{-2}$ by using data from 
high$-$ and intermediate$-$velocity gas clouds (most of them have intrinsically 
near-solar abundance). They showed that over the whole range
of HI column densities (covering almost three orders of magnitude), 
there is a tight
anticorrelation between log(X/H) and log(\N) for several elements: 
high column densities correspond to
a larger amount of depletion, with a remarkably low dispersion around each value
of \N. The slope of the anticorrelation depends on the ionic species considered;
it is $-$0.39$\pm$0.04 for MnII and $-$0.59$\pm$0.04 for FeII, two species that have
also been measured in DLAs.

It is  interesting to see whether such an anticorrelation exists for DLAs.
In Table 1 we present a compilation of currently available data for DLA
abundances that we use in this work. The redshift range of the DLAs and 
their number in each survey are also given in the Table. Data displayed 
in all our figures are directly taken from the source references, with 
no attempt to homogenize them. 

\begin{table}[h]
{\footnotesize
\noindent
Table 1. $\alpha$ and Fe$-$peak element abundances in DLAs \\ [2mm]
{\scriptsize
\begin{tabular}{ccccccccccc} 
\hline \hline 
 Ref.   & z$_{abs}$    & obj.& Al& Si & S & Cr& Mn & Fe & Ni & Zn  \\   
\hline 
(1)     & 0.3 $-$ 1.0  &  6  &   &    &   &   &    &    &    & X   \\
(2)     & 1.7 $-$ 4.4  & 15  &   & X  & X &   &    & X  &    & X   \\
(3)     & 1.9 $-$ 2.5  &  6  &   & X  & X & X &    & X  &    & X   \\
(4)     & 0.68, 1.15   &  2  &   & X  &   &   & X  & X  &    & X   \\
(5)     & 0.7 $-$ 4.5  & 23  & X & X  & X & X & X  & X  & X  & X   \\
(6)     & 2.3090       &  1  &   &    & X &   &    & X  &    &     \\
(7)     & 3.3901       &  1  &   & X  &   & X &    & X  & X  & X   \\
(8)     & 1.7 $-$ 3.0  & 15  &   &    &   & X &    &    &    & X   \\
(9)     & 0.7 $-$ 3.3  & 17  &   &    &   & X &    &    &    & X   \\
(10)    & 1.1 $-$ 1.5  &  5  &   & X  &   & X & X  & X  & X  & X   \\
(11)    & 0.6 $-$ 1.5  &  6  &   & X  &   & X & X  & X  & X  & X   \\
(12)    & 1.7 $-$ 4.2  & 22  & X & X  & X & X &    & X  & X  & X   \\
(13)    & 2.4 $-$ 4.2  &  4  &   &    &   &   &    & X  &    &     \\
\hline \hline
\end{tabular} \\ [1mm]
}
}
\noindent
{\it References:}
(1) Boiss\'e et al. 1998; (2) Centurion et al. 1998; (3) Centurion et al. 2000; 
(4) de la Varga et al. 2000; (5) Lu et al. 1996; (6) Molaro et al. 1998; 
(7) Molaro et al. 2000; (8) Pettini et al. 1994; (9) Pettini et al. 1997b; 
(10) Pettini et al. 1999; (11) Pettini et al. 2000; (12) Prochaska \& Wolfe 1999; 
(13) Prochaska \& Wolfe 2000
\end{table}

In order to evaluate the depletion in DLAs as a function
of column density, one cannot use the absolute abundances [X/H] because DLAs
have different metallicities: a low [X/H] value may correspond either to large
depletion or to little chemical evolution (or  both).
For those reasons, we adopt here the [X/Zn] ratio
as a measure of depletion of a given element X,
assuming that Zn is always undepleted, irrespectively of the
HI column density. By using [X/Zn] to measure depletion
we also assume that the real [X/Zn] value
of each DLA (i.e. before depletion) is always the same, 
which is probably not the case. However, we are looking here for systematic effects
as a function of column density; any deviations of the real [X/Zn] from the value
adopted here ([X/Zn] = 0) would merely manifest themselves as dispersion around an
average trend.

On the left panels of Fig. 1 we present the DLA data of [X/Zn] as a function of
\N  \ column density. No anticorrelation is found, contrary to the situation in gas 
clouds in our Galaxy, which is shown by {\it solid lines} in the panels of
Mn and Fe (from the fits of Wakker and Mathis 2000).
The absence of such an anticorrelation in DLAs, compared to local measurements,
seems puzzling. Does it mean that depletion (if any) in DLAs is independent of
HI column density?

\begin{figure*}
\psfig{file=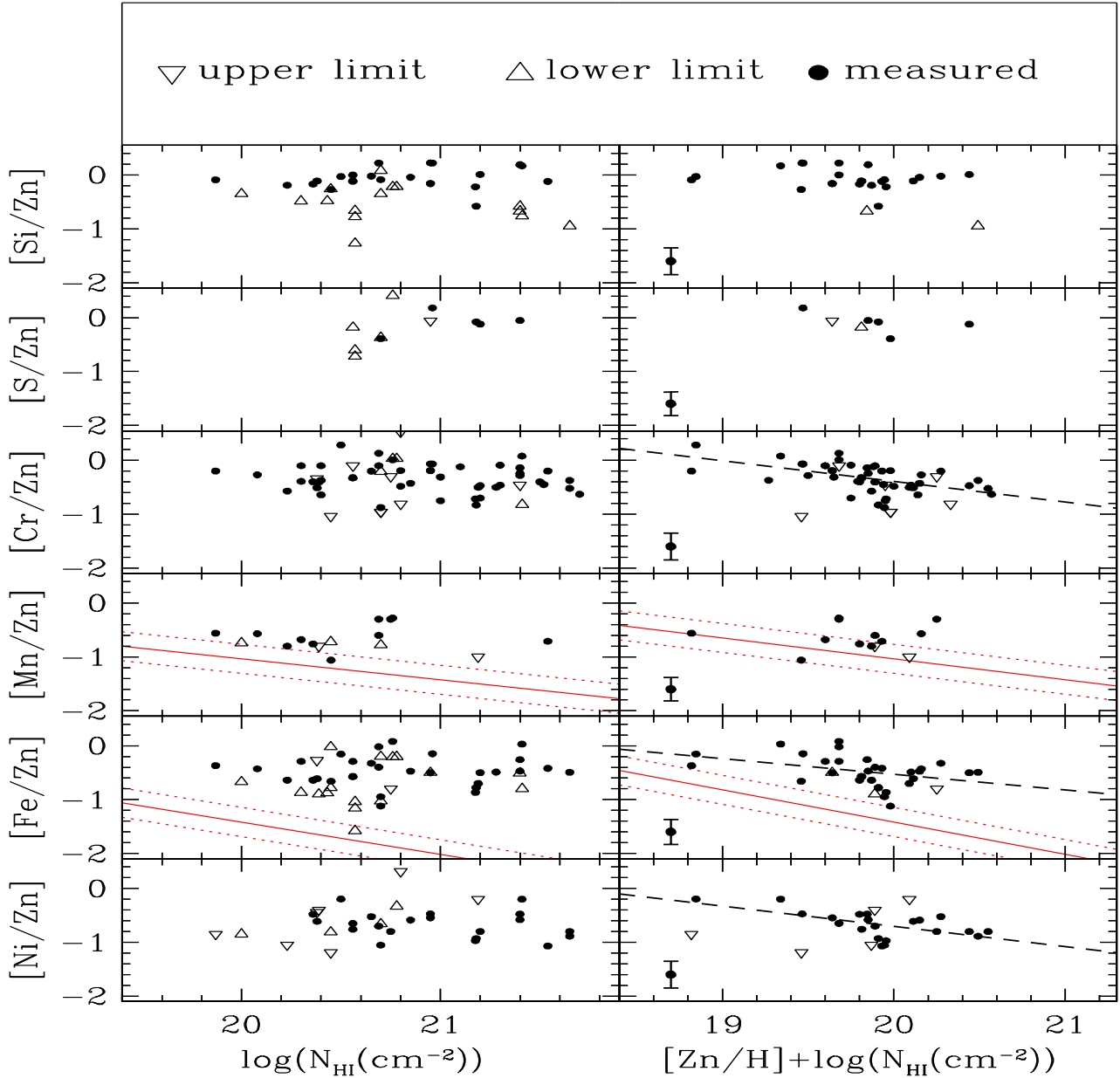,height=18.cm,width=\textwidth}
\caption{\label{}
Gas phase abundance ratios [X/Zn] in DLAs, plotted as a function of HI column
density ({\it left panels}) and of Zn column density ({\it right panels}).
Data are from references given in Table 1 and corresponding typical 
uncertainties are shown 
in each one of the right panels by {\it error bars}. 
No correlation is seen in the
left panels, while in the right panels an anticorrelation is found for the
refractory elements Cr, Fe and Ni (shown by {\it dashed lines}, obtained by a 
least squares fit to the data). In the
panels of Mn and Fe {\it solid} and {\it dotted lines} show the corresponding
trends observed in Milky Way clouds and their associated uncertainties
(see text for discussion).
}
\end{figure*}

Here, we suggest that a proper measurement of depletion is provided {\it not} by the
HI column density, but rather by the column density of (undepleted) metals, e.g. by
log(\Nz) = [Zn/H]$-$7.83+log(\N), where $-$7.83 = log(Zn/H)$_{\odot}$. In the case
of clouds in our Galaxy, metallicity is always the same 
([Zn/H] $\sim$ 0), making it possible
to use \N \ instead of \Nz. However, in DLAs, because of the different [Zn/H],
\N \  is not proportional to \Nz \ and cannot be used to measure depletion.

When [X/Zn] is plotted as a function of $\mathcal{F}$ = 
[Zn/H]+log(\N) (right panels in 
Fig. 1), one can see that: i) there is no dependence of Si/Zn  and S/Zn on
$\mathcal{F}$, as expected for those two elements which are also found not to
be substantially depleted  in local gas clouds, and ii) there 
is a clear  anticorrelation between [X/Zn] and $\mathcal{F}$ for
the refractory elements Cr, Fe and Ni (as can be seen from the {\it dashed lines}) 
while the scarce data for Mn do not allow  for a conclusion.

We think that the anticorrelation obtained between [X/Zn] and  $\mathcal{F}$ in DLAs
is an unmistakable sign of depletion for Fe, Cr and Ni. Metal column density is
the appropriate factor, since it measures the total amount of metals along the line
of sight, and thus the amount of dust; \N \  alone is inappropriate, since 
further assumptions about the dust to gas ratio have to be made.

How can one explain the fact that [X/Zn] seems to be independent of \N \ for all 
elements in DLAs? We note that there exists
a well known anticorrelation between metallicity, as expressed by [Zn/H], and
HI column density in DLAs. This anticorrelation, first noticed by Boiss\'e et al. 
(1998), is attributed to  selection biases (since Zn is not depleted into 
dust grains): regions of high [Zn/H] and high \N \ are opaque to light from
background quasars and thus undetectable, while regions of low [Zn/H] and low \N 
(i.e. of low Zn column density) are not detectable by current DLA surveys. In PB2000a
we substantiated this argument by running detailed chemical evolution models for 
disk galaxies. We  showed how this ``filter'' can lead to a ``no-evolution'' 
picture for the
average DLA metallicity as a function of redshift, in agreement with observations.
We shall return to that question in Sec. 3.3, presenting results for other elements
than Zn. Here we simply note that, since high \N \ DLAs are asociated, on average,
with low [Zn/H] and vice$-$versa, the depleted ratio [X/Zn] is expected to be, on
average, independent of \N, as observed. On the contrary, when [X/Zn] is plotted
as a function of  $\mathcal{F}$ (which is directly proportional to the amount
of dust), the anticorrelation expected by local observations becomes clear.

An inspection of the Fe panel in Fig. 1 shows that the absolute 
[Fe/Zn] values are systematically higher in DLAs than in the Milky Way clouds 
(by $\sim$ 0.7 dex);  a similar remark also holds for 
Mn, although the scarcity of data makes its case less compelling.
If metal column density is the
relevant factor, as argued in the previous paragraphs, how can this systematic
offset (by factor of $\sim$ 5) be understood? We see two possibilities:

1) The undepleted [Fe/Zn] in DLAs is not zero (i.e. the Fe/Zn ratio is
not solar) as assumed here, but
$\sim$ 0.7 (i.e. 5 times solar); however, despite current uncertainties on the
nucleosynthesis of Zn, this possibility is quite implausible. Indeed, Fe/Zn is
solar on average in both halo and disk stars of the Milky Way.

2) Wakker and Mathis (2000) suggest that the very small scatter around the average
depletion pattern in the Milky Way clouds (within {\it dotted lines} in the Fe 
panels of Fig. 1) is due to the fact that one single cloud dominates the column 
density of each line of sight. In that case, a single value of (depleted) Fe/Zn 
corresponds to a given value of \N \ (or of  $\mathcal{F}$, since [Zn/H] $\sim$ 
solar in the Milky Way clouds). In the case of DLAs, it is tempting to assume 
that a given line of sight intercepts several clouds: the depletion of [Fe/Zn] 
within each cloud would be small, but the total column density of metals 
($\mathcal{F}$) would be high. This picture could explain the offset of DLAs 
with respect to the local pattern in Fig. 1. Taking into account the error bars 
in [Fe/Zn] values and the possibility of systematic errors, it is difficult 
to tell whether a couple of clouds or a larger number is necessary.

 At this point, we notice that in the case that DLAs are composed by many
  systems of individual column densities logN(HI)$<$19.5, ionisation effects
  could play an important role in shaping the observed abundance ratios. As
  suggested recently by Howk and Sembach (1999), the observed ratios may
  systematically overestimate the intrinsic ones in the cases of Si/Fe and 
  Zn/Cr, if ionised gas is present in these systems, 
  "thereby mimicking the effects of alpha-element enrichment or
  dust depletion". Izotov et al. (2000) have further explored this issue, on 
  the basis of similarities observed between some DLAs and nearby blue compact
  dwarf galaxies; using a schematic two zone model, they found that ionisation
  effects may play some role in shaping the observed 
  abundances of these systems, although they recognised that their model
  suffered from various uncertainties related to the unknown structure and
  geometry of the absorbing systems, the spatial distribution of the ionising
  sources and their radiation field.
  We do not consider any further this possibility here, since
  we think that the anticorrelation displayed in Fig. 1, i.e. the  
  dependence of X/Zn on Zn column density for well known refractory elements 
  (like Fe, Ni and Cr) but not for e.g. Si,  suggests that dust depletion is 
  the main factor.
 
The observations of Fig. 1 tell nothing about the real (undepleted) abundances in
DLAs; further assumptions have to be made for that.
The investigation of the abundances and the depletion patterns in DLAs by means
of a consistent chemical evolution model is the subject of the following sections.

\section{ DLAs as disk galaxies}

The details of our model for the chemical evolution of spiral disks are
described in Boissier and Prantzos (2000, 1999; hereafter BP2000 and BP1999, 
respectively),  while its application to the evolution of Zn in 
DLAs is made in  Prantzos and Boissier (2000a, hereafter PB2000a). 
In Sect. 3.1 we briefly recall the main features of the model and its  success 
in reproducing a large number of observations concerning present day disks.
In Sect. 3.2  we present in some detail the only two novel 
ingredients with respect to BP2000 and BP1999, namely the metallicity 
dependent yields of Woosley and Weaver (1995, hereafter WW1995) for 
intermediate mass elements, and the new yields of SNIa from  
of Iwamoto et al. (1999). In Sec. 3.3
we extend our study  to the evolution of intermediate mass elements observed 
in DLAs.

\subsection{Description of the model} 

A detailed model of the chemical and spectro-photometric evolution
of the Milky Way is constructed in BP1999. The disk 
of our Galaxy is simulated as an ensemble of concentric rings gradually 
built up by infall of gas of primordial composition. We adopted
a Star Formation Rate (SFR) suggested by the theory of density waves 
$\psi \propto \Sigma_{gas}^{1.5} V(R)/R$ (where $V(R)$ is the circular velocity 
at radius $R$), which is in agreement with the observed current SFR profile 
in the Milky Way.
The infall rate is exponentially declining, 
with a timescale $\tau$ = 7 Gyr in the local disk
as to reproduce the observed G-dwarf metallicity distribution. 
Inner galactic zones are formed earlier than the outer disk 
(formation ``inside-out'').
The number of observables in our Galaxy, concerning the history of the
solar neighborhood as well as the current profiles of various quantities,
is larger than the few  ``free'' parameters involved in this model and testifies 
to its validity.

After ``calibrating'' the model on the Milky Way (in fact, the SFR efficiency and
the infall timescales as a function of local surface densities)
we extended it to the study of local spirals (BP2000). 
We used simple ``scaling laws'', suggested by Mo, Mao and White (1998) 
in the framework of  Cold Dark Matter scenarios of galaxy formation. 
Within this simplified approach, disks are characterised by
two parameters: circular velocity $V_C$ 
(determined by the mass of the dark halo), 
and spin parameter $\lambda$ (measuring the specific angular momentum
of the dark halo). For instance,
the disk scalelength is $R_d = R_{d,MW} * (V_C/V_{C,MW}) * (\lambda/\lambda_{MW})$,
where the index $MW$  refers to the corresponding value in the Milky Way
(see BP2000 for details).

Our grid of models covers the range 80 $<V_C {\rm (km/s)} <$ 360
and 1/3 $< \lambda/\lambda_{MW} <$ 3, which corresponds to current High Surface
Brightness disks. Our results reproduce quite satisfactorily
a large number of observed properties of present-day disks: sizes, 
surface brightness, Tully-Fisher relationship, colours, abundances, 
integrated spectra, etc... (BP2000). Also, the model reproduces the 
observed abundance gradients in disk galaxies (Prantzos and Boissier, 2000b)
and the gas fraction, star formation efficiency and effective ages of a large sample
of nearby spirals (Boissier et al. 2000). We notice that, according to our models,
massive disks form quite rapidly (in a few Gyr) while it takes several Gyr for the
small ones to be formed.

In order to apply the model to the study of DLAs, the following statistical factors were
taken into account:

First, at redshift $z$ = 0, a velocity distribution function can be obtained by 
combining a Schechter luminosity function and an observed Tully-Fisher relationship, 
as in Gonzalez et al. (2000):
\begin{equation}
F_V(V) dV = \tilde{\Psi}_* \left(\frac{V}{V_*} 
\right)^{\beta}exp\left[-\left(\frac{V}{V_*}\right)^n\right]\frac{dV}{V_*}.
\end{equation}
For the parameters of Equ. (1) we adopted the values
of the fifth row of Table 4 in Gonzalez et al. (2000), corresponding
to the velocity interval covered by our models;  another choice would
have small impact on the results of this paper.

Results of numerical simulations 
(see e.g. Mo, Mao and White, 1998)
give the distribution function of the spin parameter $\lambda$:
\begin{equation}
F_{\lambda}(\lambda)d\lambda \ = \ \frac{1}{\sqrt{2\pi}\sigma_{\lambda}} \
exp \left[ - \frac{ln^2(\lambda/\bar{\lambda})}{2 \sigma_{\lambda}^2} 
\right]\frac{d\lambda}{\lambda}
\end{equation}
with $\bar{\lambda}$ = 0.05 and $\sigma_{\lambda}$ = 0.5.

Finally, the probability that a line of sight to a QSO intercepts a disc
in the radius interval $[R,R+dR]$ is proportional to the geometrical
cross-section $F_R(R) dR =  2 \pi R dR/ \pi R_M^2$ (where $R_M$ is the radius 
of the  largest disc in our models),
favouring the detection of the outer regions of the larger disks.

At any time, or redshift, the average value of a quantity Q 
(for instance an abundance ratio) in our models
is computed from:
\begin{equation}
\label{EQUAaver}
\left< Q \right> = \frac{\int_{\lambda} \int_{V_C} \int_{R} 
F \Phi(\mathcal{F}) Q(\lambda,V_C,R) \ dR \ dV_C \ d\lambda}
{\int_{\lambda}\int_{V_C}\int_{R} F \Phi(\mathcal{F}) \ dR \ dV_C \ d\lambda }
\end{equation}
where $F = F(\lambda,V_C,R)=F_V(V_C) F_{\lambda}(\lambda) F_R(R)$.
We implicitly assume that those three functions are mutually independent
and also time-invariable. This is of course an over-simplification,
valid as long as mergers and  intereactions do not modify
too much $F_V$ and $F_{\lambda}$. This is probably a good approximation
at low redshifts, but becomes more and more problematic at
high redshifts ($z>3$).

The function $\Phi$ in Equ. (3) plays the role of a ``filter''.
To obtain the true mean value over the whole ensemble of disks, 
one should take  $\Phi(\mathcal{F})$ = 1.
However, Boiss\'e et al. (1998) argued that observations of DLAs suffer from
selection  biases, as discussed in Sec. 2.
Measurement of zinc abundances vs. HI column density show an anti-correlation,
resulting from a bias against too large or too small values of 
$\mathcal{F}$ = [Zn/H]+log(N(HI)).

To account for that bias, we apply Equ. (3)
with $\Phi(\mathcal{F})$ = 1 for $18.8<\mathcal{F}<21$  and 0 elsewhere; this
gives the average value of $Q$ over those parts of our disks that have
DLA-like properties (at least, as far as Zn abundances and HI column 
densities are concerned). 
As shown in PB2000a, this empirical ``filter'' can explain 
the weak evolution with redshift of [Zn/H], and the corresponding average value
and observed scatter among DLAs
(see also top-left pannel in Fig. 2 and discussion in Sec. 3.3). 
In the present work we extend the model to the study of other elements;
the adopted yields are presented in the next section.

\subsection { Yields of massive stars and SNIa }

An important ingredient in studies of galactic
chemical evolution is the stellar  yields of various elements. 
Massive stars are the main producers of most heavy 
isotopes in the Universe. Elements up to Ca are mostly produced in such 
stars by hydrostatic burning, whereas iron peak elements are produced 
by the final supernova explosion (SNII), and by white dwarfs 
exploding in binary systems as SNIa. Most of He, C, N and minor CO 
isotopes, as well as s-nuclei come from intermediate mass stars(2$-$8\ms).
Since we are mainly concerned with the abundances of $\alpha$ and 
iron-peak elements in DLAs, we consider no yields from intermediate 
mass stars in this work. 

We use the metallicity dependent yields of WW1995, which are given for stars 
of mass M = 12, 13, 15, 18, 20, 22, 25, 30, 35 and 40 \ms and metallicities
Z/\zs = 0, 10$^{-4}$, 10$^{-2}$, 10$^{-1}$ and 1. 
The WW1995 yields, folded with a stellar Initial Mass Function (IMF)
lead to approximately solar abundance ratio of O/Fe
(or $\alpha$ element), without the Fe contribution of SNIa.
This lead Timmes et al. (1995b) to suggest that the Fe 
yields of WW1995 are probably overestimated. Following their suggestion,
we adopted in our model half the nominal 
values for the WW1995 yields for iron peak 
elements (from Cr$-$Zn). Taking into account the uncertainties currently 
affecting those yields, such a reduction is not unreasonable. 
Our procedure allows to produce the observed evolution of O/Fe and $\alpha$/Fe
in the Milky Way halo, 
and does not alter the abundance ratios between iron peak elements (which 
are most commonly observed in DLA systems). 

To account for the additional source of Fe-peak elements, required to 
explain the observed decline of O/Fe abundance ratio in the Milky Way disk 
(e.g. GP2000), we utilise the recent yields of SNIa 
from the exploding Chandrashekhar-mass CO white dwarf models W7 and W70
of Iwamoto et al. (1999). These are updated versions of the original
W7 model of Thielemann et al. (1986), calculated for metallicities
Z = \zs (W7) and Z = 0 (W70), respectively. 
These SNIa models lead to an oveproduction 
of Ni, which will also be clearly seen when we compare our model 
results of [Ni/Zn] with observations in low redshift DLAs. 

Finaly, notice that, due to  uncertainties in the Zn yields of WW1995
(see Timmes et al. 1995b and PB2000a), we assume that Zn behaves always as Fe
and we use then our model values of [Fe/H] as predictions for [Zn/H]. 
This is justified by the fact that observations in the Milky Way disk 
and halo show [Fe/Zn] = 0 for a large range of metallicities. However, we shall
question that assumption in Sec. 4.2.

In recent papers we have shown how these metallicity dependent yields, combined
with the model described in Sec. 3.1, lead to satisfactory results concerning the
chemical history of the solar neighborhood (GP2000) and the current
abundance gradients of intermediate mass elements in the the disk of the Galaxy
(Hou et al. 2000). The use of metallicity dependent yields is a necessary
condition for a consistent study of the abundance patterns in DLAs.

\subsection { Metallicity evolution vs observations}

We calculated the evolution of the abundances of all isotopes between H and Zn
in the different zones of our disk models. In this section we present results for
eight elements (Zn, Al, Si, S, Mn, Cr, Fe and Ni) that have been observed in
DLAs.

In Fig. 2 we present the absolute abundances of each element as a function of redshift.
We display our results in two different ways. First, we show in {\it long-dashed
curves} the extreme values, i.e. the highest and lowest metallicities at a given 
redshift, obviously corresponding to the innermost zones of massive disks and the
outermost zones of low mass disks, respectively (defined in PB2000a and 
here to be the zones at radial distance of 0.5 $R_d$ and 5.5 $R_d$, respectively). 
The corresponding average value, taking into account various statistical
factors (as discussed in Sec. 3.1) is
given by Equ. 3 with $\Phi(\mathcal{F})$ = 1 in all zones and is shown by a 
{\it dotted} curve.
It can be seen that for all elements  this ``unfiltered'' average value, 
as well as the extreme  values, display strong evolution and are in disagreement 
with the data.

The {\it shaded area} of Fig. 2 is obtained by adopting the ``filter'' suggested
by Boiss\'e et al. (1998) for Zn and already introduced in PB2000a, 
i.e. by putting in Equ. 
(3) $\Phi(\mathcal{F})$ = 1 for 18.8 $ <\mathcal{F} <$ 21, and $\Phi(\mathcal{F})$ = 0
otherwise. The upper value of this filter ($\mathcal{F}<$21) is maintained
for all elements, since for larger values, obscuration of background quasars
inhibits DLA detection (at least according to our interpretation of the
observed Zn/H vs \N \  anticorrelation, see PB2000a). However, the lower limit
($\mathcal{F}>$ 18) adopted for Zn corresponds to the minimal column density
of Zn atoms detectable in current surveys; it should not be the same for other 
elements, since it depends (among other) on which ionisation lines are used 
for abundance determination. We adopt then as a conservative lower limit 
for all other elements the
condition log(\N) $>$ 20.2, which is simply the definition of a DLA.
In other terms, we consider only the zones of our models that have surface
densities $\Sigma_{Gas}>$2 M$_{\odot}$ pc$^{-2}$.

Applying this filter to our results (i.e. by excluding the very-metal rich
and dense inner zones, as well as the low-density metal poor zones) we find 
a much narrower range of abundances at a given redshift, as well as a milder
abundance evolution than in the original models
(the {\it solid} curve in Fig. 2 gives the average value in the ``filtered''
zones, i.e. those with log(\N)$>$ 20.2 and $ \mathcal{F} <$ 21).
In PB2000a we found that this
``biased evolution'' picture is quite compatible with observations of Zn in
DLAs. Here we show that it is also compatible with all available data on
DLA abundances: indeed, both the observed dispersion at a given redshift 
and the mild overall evolution are nicely reproduced.

\begin{figure}
\psfig{file=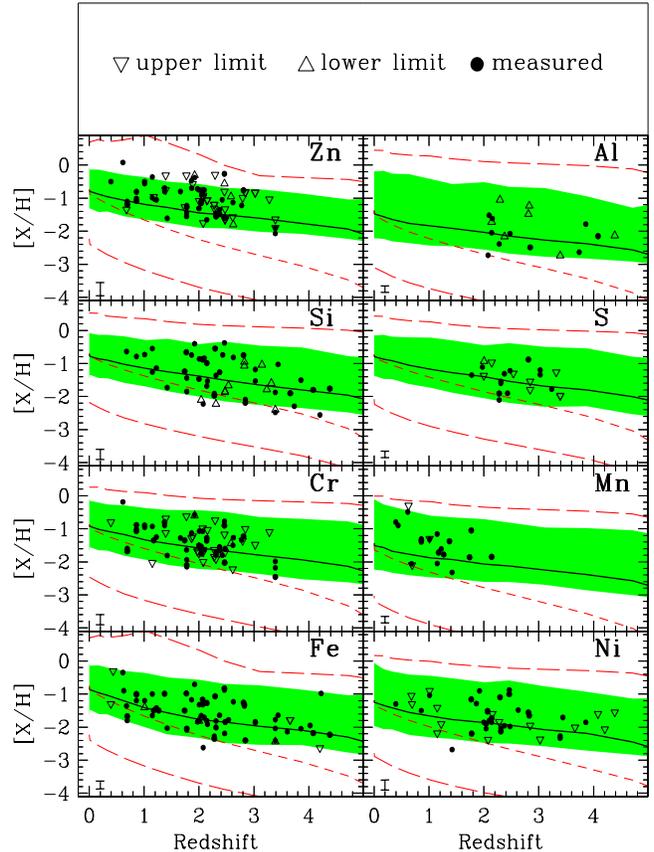,height=13.cm,width=0.5\textwidth}
\caption{\label{}
Evolution of elemental abundances in our disk galaxy models for 8 elements,
Al, Si, S, Cr, Mn, Fe, Ni and Zn, as a function of redshift $z$. Observational data
for DLAs are from references listed in Table 1, with
typical errors shown in the bottom left corner of each panel.
In each panel, the {\it shaded area} includes 
all the ``filtered'' zones of our models (i.e. by taking various biases
into account, see text) and the {\it thick curve}  
is the corresponding mean value in those same zones.
A weak evolution and an important scatter, both compatible with
observations, are obtained for all elements. 
For comparison, we also give the results corresponding to the 
``unfiltered situation'' (i.e. all the zones of our models): 
the {\it long dashed } curves correspond the upper and lower limits, while 
the {\it short dashed } curve is the corresponding mean value.
Our disk models (which include no correction for dust depletion
effects) have a true metallicity evolution stronger than the
one observed in DLAs.
}
\end{figure}

We find that in our models the average abundance  varies 
at a rate  $d<[\alpha/H]>/dz$ $\sim$ $-$0.25 for $\alpha$$-$elements and 
$d<[Fe/H]>/dz$ $\sim$ $-$0.33 for Fe in the redshift range $z$ = 0.5$-$4. 
In the latter case, the larger rate is due to the introduction of a 
late source of Fe, namely SNIa. We note that in a recent work
Savaglio e al. (1999), after correcting for depletion a large number of
adundances in DLAs find that the evolution of the intrinsic DLA metallicity
Z as a function of redshift can be described by dlogZ/d$z$=-0.30$\pm$0.06
for 0.5$<z<$4, i.e. a value not very different from ours.

The scatter of absolute abundances at a given redshift in our models
is attributed both to the different types of galactic disks considered
here (massive ones evolving more rapidly than low mass ones) and to the
differential evolution of the various regions within a given disk
(inner zones evolving more rapidly than the outer ones).
The obtained scatter, typically a factor of 30$-$50 at a given redshift,
reproduces fairly well the observed one.
We notice that in Fig. 2 we find a smaller scatter in the ``filtered''
values of Zn than in those of Fe, although we assumed that 
both elements evolve in the same way (i.e. [Zn/Fe] = 0). This is due to the
different lower filters used: $\mathcal{F}>$ 18.8 for Zn and log(N(HI))$>$20.2
for Fe.

The mild evolution and the large corresponding scatter of metal abundances
in DLAs have been addressed in previous works in the field. Observational
papers attribute the former to selection bias (e.g. Vladilo et al. 2000)
although no quantitative model is proposed to substantiate this argument.
On the other hand, most theoretical papers (e.g. Timmes et al. 1996,
Ferrini et al. 1997, Prantzos and Silk 1998, Jimenez et al. 1999, 
Lindner et al. 1999) ignore such biases and find invariably a metallicity 
evolution stronger than observed. Selection biases are properly considered
in PB2000a. Meusinger and Thon (1999) also take such effects into account,
but consider only disk galaxies in a narrow range of masses
(5$-$8 10$^{10}$ \ms), i.e. with quite similar star formation histories.
As a result, even after considering selection effects, they find that
the average metallicity in their models increases by a factor of $\sim$ 10
between $z$ = 1 and 3, much larger than suggested by the observations.
For that same reason, they fail to reproduce the observed scatter at a 
given redshift and they obtain much smaller values.

\section{Abundance ratios and depletion patterns in DLAs}

Abundance ratios are more reliable tracers of the chemical evolution than
absolute abundances. They allow to identify whether an element originated
from long lived sources (low mass stars or SNIa) or short lived ones 
(massive stars and SNII), or whether it is affected by the odd-even effect. 
The analysis of abundance ratios in the Milky Way disk and halo stars does
provide important insight on the chemical history of the Milky Way 
as well as on the different  production sites of various elements
(despite the existence of large uncertainties in both theory and 
observations, see GP2000). 

However, situations in DLAs are more complicated. The abundances 
observed in DLAs may not represent the real chemical composition of the 
system if part of the elements is removed from the gas to the solid phase 
(dust grains), as happens in the interstellar medium of our Galaxy. 
In the discussion of Sec. 3.3, we did not consider any such kind of dust
depletion effects. We have argued, in Sec. 2, that the observed abundance 
ratios [X/Zn] in DLAs, when plotted as a function of  $\mathcal{F}$, 
suggest that some depletion does occur in those systems, at least for 
Fe, Cr, Mn and Ni. We have also argued that the depletion pattern for Fe, 
when compared to the one observed in the Milky Way, suggests that the 
line of sight intercepts several gas clouds in a given DLA.

In this section, we compare observed abundance patterns in DLAs with 
those obtained in our models of disk galaxy evolution. Our
aim is to find out : i) whether there are any discrepancies between
models and observations, ii) whether such discrepancies can be
attributed to depletion, and iii) whether the resulting depletion
patterns correspond to well known ones, such as those observed in
the local disk or  halo clouds.

\subsection{Abundance ratios in DLAs} 

In Fig. 3 we plot the evolution of abundance ratios [X/Zn] as a function 
of metallicity ( = [Zn/H]), redshift and metal column density ( $\mathcal{F}$ ) 
for the  elements Al, Si, S, Cr, Mn, Fe and Ni. As in many other works, 
we assume that Zn is not depleted at all in dust and, therefore, 
it can be used as a tracer of metallicity. As explained in Sec. 3.2, 
because of uncertainties  about the nucleosynthesis of Zn, we assume that  
[Fe/Zn] = 0 always in our model.
The {\it heavy shaded} regions in Fig. 3 present the results corresponding
to the ``filtered'' zones of our models (i.e. satisfying  DLA constraints),
while the {\it light shaded} ones correspond to all other zones.

\begin{figure*}
\psfig{file=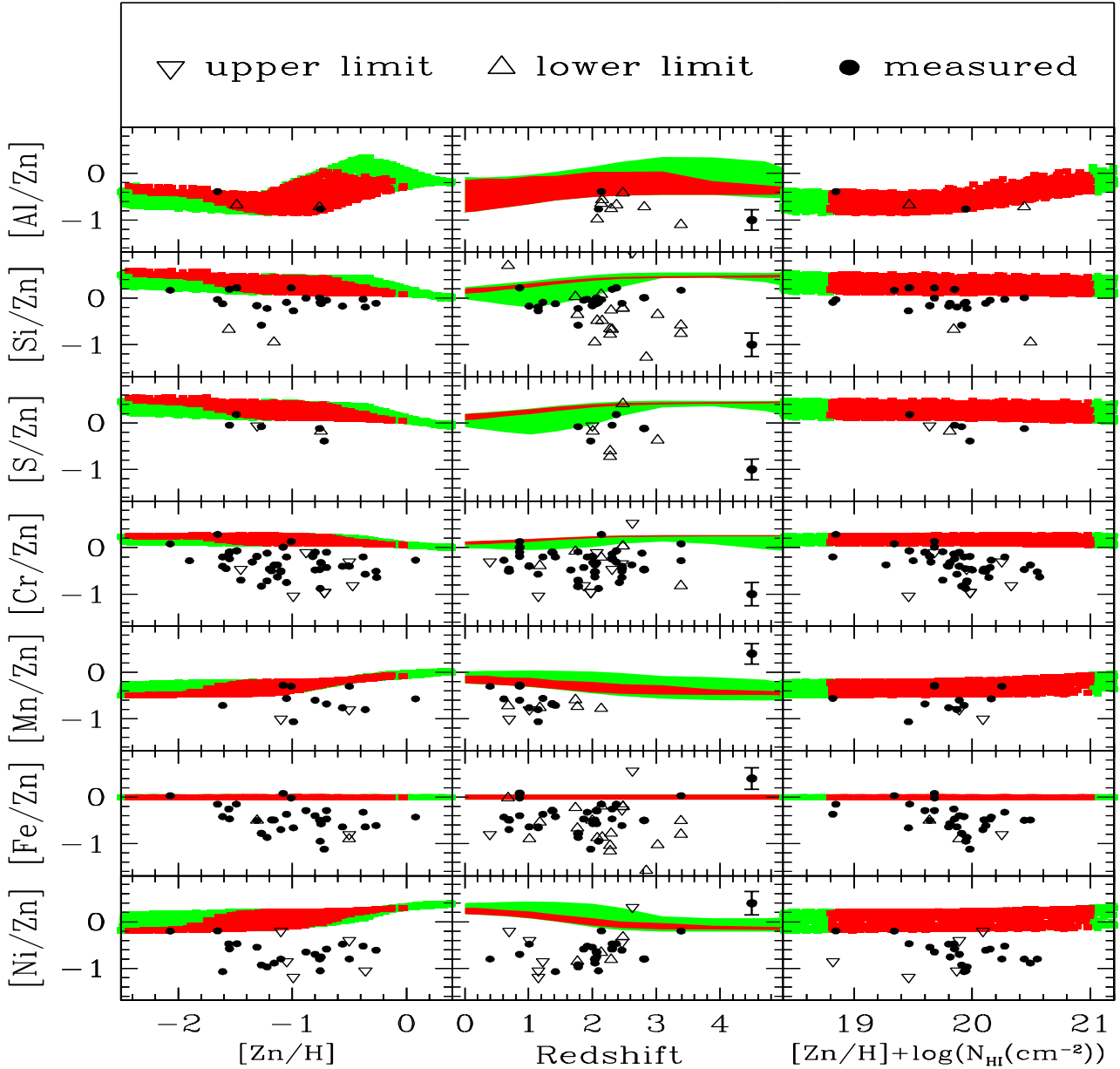,height=18.cm,width=\textwidth}
\caption{\label {figXZn} 
Abundance ratios versus metallicity ({\it left}) redshift ({\it middle}) 
and metal column density $\mathcal{F}$ ({\it right}). In all panels, 
the {\it light shaded area} includes all the zones of our models, 
while the {\it heavy shaded area} only those corresponding to the 
``filtered'' zones (i.e. satisfying DLA constraints, see Sec. 3.3).
Note that [Fe/Zn] = 0 in our models, because we have assumed that 
[Zn/H] = [Fe/H] always (see Sec. 3.2). 
Data for DLAs are from Table 1 and typical error bars are displayed 
in the middle panels. Deviation of the data from the model values 
{\it at a given redshift} is, in principle, attributed to  dust depletion,
which is not included in the models (see discussion in Sec. 4.2). 
}
\end{figure*}

{\sc $\alpha$ elements} Si and S: $\alpha$ elements present a well known 
behaviour in our Milky Way disk and halo.  The [$\alpha$/Fe] ratio is
nearly constant in the halo, at a value of 0.3 $\sim$ 0.5, and declines 
gradually in the disk from [Fe/H] $\sim$ $-$1 to 0. The later feature is 
attributed to the late ($\sim$ 1 Gyr) contribution of SNIa to the disk 
composition. 

The observed [Si/Zn] ratios in DLAs do not show any trend with metallicity,
redshift or column density, and they are solar, on average (taking into
account observational uncertainties). Our model results, when plotted
as a function of metallicity or column density cannot be directly compared
with the corresponding data, because a given metallicity in the model
may be reached at widely different epochs, dependent on the chemical
history of the corresponding region. Only when [X/Zn] is plotted as a
function of redshift can a meaningful comparison to the data be made.
From the middle panels of Fig. 3 it can be seen that at early times
($z>$ 2.5) the model [Si/Zn] $\sim$ 0.4, with a very small variation around
that value.

At late times [Si/Zn] declines in general, albeit with large variations. 
The decline is due to the late appearence of Fe producing SNIa and to our
assumption that [Fe/Zn] = 0 always. There are large variations of [Si/Zn]
between the various zones of our models ({\it light shaded area}), 
because SNIa enter the scene at 
different times depending on the previous SFR history of the corresponding
zones. In the ``filtered'' zones, however, the scatter around the mean value 
is very small. Compared to that well defined model trend, the data show 
systematically lower values. This difference could, in principle, be attributed
to depletion by $\sim$ 0.4 dex, an amount comparable with the one seen
in the local warm clouds of the disk and the halo.

However, [S/Zn] in our models behaves exactly as [Si/Zn] (for the same reasons)
and the corresponding DLA data suggest again the same degree of depletion. This
is difficult to understand, since S is not expected to be depleted at all.
Ockam's razor, along with the fact that S/Zn and Si/Zn in DLAs are independent
of Zn column density (Fig. 1) suggest then that these two elements are 
not depleted in DLAs; they also suggest that the intrinsic DLA ratios
of those elements are $\sim$ solar, irrespectively of redshift.

This conclusion does not answer the question about the nature of DLAs.
Several authors (Centurion et al. 2000; Jimenez et al. 1999, Vladilo 1999)
suggested that Low Surface Brightness galaxies (LSBs) are good candidates,
since they probably have different nucleosynthetic histories (at present 
unknown and unconstrained). At this point we emphasize that our scenario
involves many different chemical histories: disks of different masses 
form the bulk of their stars at different epochs and within them different
zones evolve at different rates. As a result, at late times, the [$\alpha$/Zn]
ratio varies considerably in our models. But the zones satisfying the DLA
criteria ( $\mathcal{F}>$ 18.8 for Zn and log(N(HI))$>$ 20.2 for other elements)
display a narrow range of  values at all times.
We shall return to that point in Sec. 4.2.

{\sc {\rm Fe} peak Elements}: Cr, Mn, Fe and Ni are produced by massive stars
in the early life of the Milky Way and mostly by SNIa at late times.
Observations of their abundances in halo and disk stars show that Cr, Fe and
Ni behave similarly, at least down to [Fe/H]$\sim$ $-$2, while the monoisotopic
Mn behaves differently: the [Mn/Fe] ratio is $\sim$-0.4  in the halo and
increases to its solar value in the disk. This behaviour is well accounted 
for by the metallicity dependent yields of WW1995 for massive stars and those
of Iwamoto et al. (1999) for SNIa, as shown by detailed chemical evolution
models in GP2000.

Our models show that the [Cr/Zn] and (by assumption) [Fe/Zn] ratios
are $\sim$ solar and independent of metallicity or redshift. As expected, 
[Mn/Fe]$\sim$ $-$0.4 at low metallicities (high redshifts) and increases 
slowly at late epochs. The small overproduction obtained for [Ni/Zn] obtained
at late times has a well known origin: the adopted SNIa yields of Iwamoto
et al. (1999) which oveproduce Ni by a factor of $\sim$ 2 (see GP2000
for a discussion). Taking [Ni/Zn] = 0
over the whole metallicity range would thus be more appropriate.

In all four cases, the observed [X/Zn] ratio in DLAs is clearly below solar or
model values, pointing to depletion as the most obvious explanation. This
conclusion is corroborated by the observed anticorrelation of [X/Zn]
with Zn column density, as argued in Sec. 2.

In a recent work Pettini et al. (2000) selected a small subsample of [Mn/Fe]
values in DLAs with [Zn/Cr] $<$ 0.3, i.e. with small amount of depletion; 
assuming then that all Fe peak elements are depleted by the same (small) amount,
they found that [Mn/Fe]$\sim$ $-$0.4 in DLAs, independent of metallicity.
This picture is not quite compatible with the one in the Milky Way, but because
of their small sample it is difficult to conclude whether their finding 
has implications for the role of SNIa in DLAs or not.

\subsection{Depletion patterns} 

\begin{figure*}
\psfig{file=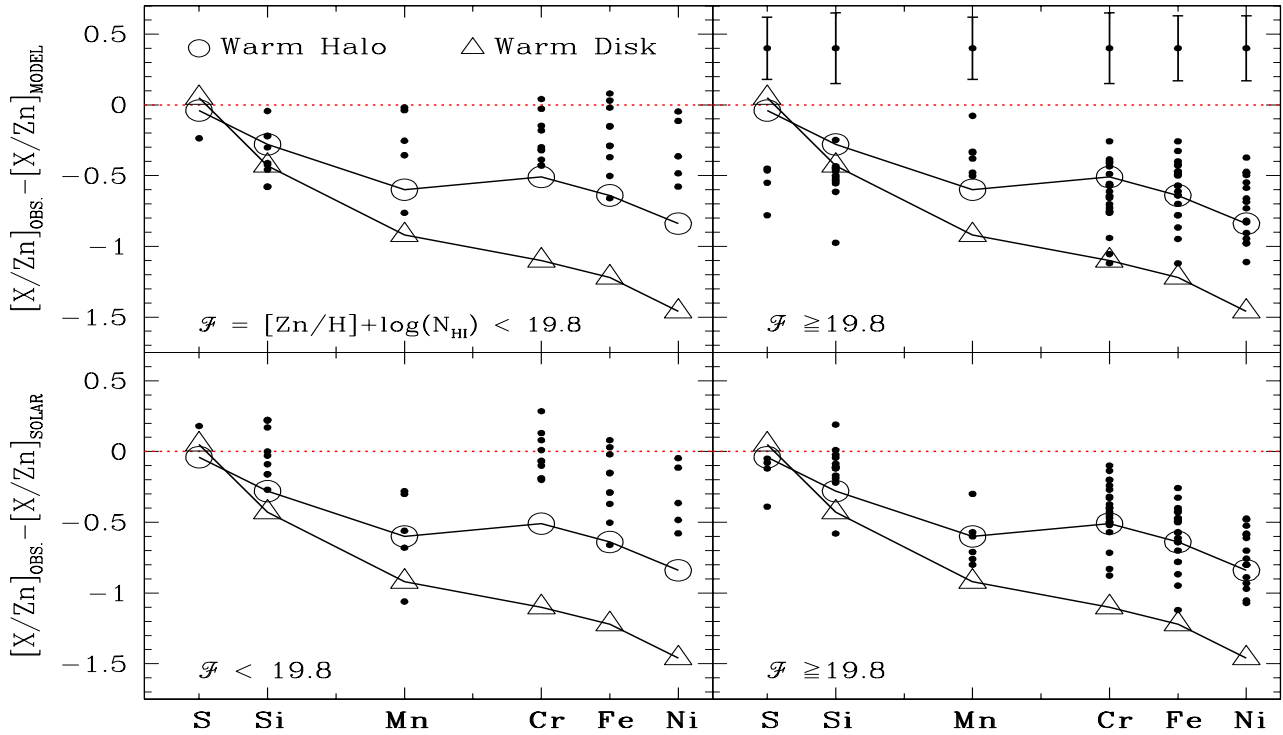,angle=-90,height=11.cm,width=\textwidth}
\caption{\label{} 
Depletion patterns in DLAs. In the {\it upper panels}, observed DLA
abundance ratios [X/Zn] are compared to those resulting from 
our models, {\it taken at the absorber's redshift}.
In the {\it lower panels}, the intrinsic DLA abundance pattern is
simply assumed to be solar. In both cases, the DLA data are plotted
according to the corresponding Zn column density, given by
$\mathcal{F}$ = [Zn/H]+log(\N): {\it left} panels correspond to 
$\mathcal{F}<$ 19.8  and {\it right} panels to $\mathcal{F} \ge$ 19.8.
In all panels the depletion patterns obtained in warm halo and
disk clouds in the Milky Way are shown by {\it open symbols},
connected with solid lines (data from Savage and Sembach 1996). 
Typical error bars for measured
[X/Zn] ratios in DLAs are displayed in the {\it right upper panel}.
}
\end{figure*}

From the discussion of the previous section it turns out that several elements
in DLAs are depleted, at least with respect to the predictions of our 
chemical evolution models, which reproduce satisfactory several DLA features
(e.g. Fig. 2). It is interesting to see how the predicted depletion patterns
compare to some well known ones, as those of the warm halo and disk clouds
in the Milky Way.

In the upper panels of Fig. 4, we compare the observed [X/Zn] ratios in DLAs
to the corresponding ones of our models {i.e. \it at the  absorbers' redshifts},
(as given in the middle panels of Fig. 3). Taking into account the discussion 
of Sec. 2, namely the fact that the depletion patterns depend on Zn column 
density, we split the data into low and high $\mathcal{F}$ values 
($\mathcal{F<}$ 19.8 and  $\mathcal{F} \ge$ 19.8, respectively). We also show, 
for comparison (in both panels) the depletion patterns in the Milky Way warm 
halo and disk clouds ({\it open symbols}, connected by solid lines, data 
from Savage \& Sembach 1996).
Due to the uncertainties (unknown intrinsic abundances in DLAs, measurement
errors), such a comparison has no sense for individual DLA, but only in
a statistical sense.

In the left upper panel of Fig. 4, it is seen that the depletion amount
obtained at low $\mathcal{F}$ is, on average, smaller than in warm halo clouds.
The average depletion is similar for all elements, $\sim$ $-$0.2 dex, and comparable
to the error bars of the measurements. In the right upper panel of Fig. 4, the
depletion is slightly larger for all elements. The depletion pattern ressembles,
on average, the one of warm halo clouds for Fe, Ni and Cr, but also for Mn and Si
(i.e. within error bars). However, as dicussed in the previous section, it is 
simply impossible to accomodate S in this scheme, since it is hard to
imagine that its ratio to Zn can be subsolar. 

The problem with S may be due to the fact that we use as reference pattern the one 
resulting from our models. We checked then whether using the solar abundance ratios
as reference pattern, could lead to a better fit of the data. The results of that
comparison appear in the lower panels of Fig. 4, again for low and high 
$\mathcal{F}$ values (left and right lower panels, respectively).
At low $\mathcal{F}$ hardly any depletion at all is found, for all elements except
Mn; this is puzzling, since Mn can be no more depleted than the other Fe peak
elements. At high $\mathcal{F}$, the resulting depletion pattern compares
fairly well with the one of the warm halo clouds for all elements.

In summary: if one assumes that the intrinsic metallicity pattern in DLAs is
the one given by our models, a difficulty arises with S which is found to be
depleted at high $\mathcal{F}$. If on the other hand it is assumed that the
intrinsic DLA pattern is solar, a problem arises with Mn, which is found to be
depleted at low $\mathcal{F}$.

We note that our assumption of [Fe/Zn]=0 always, although based on 
observations of Milky Way stars,
may be inconsistent and potentially wrong.
Indeed, it implies that $\sim$2/3 of Zn are produced {\it at late times} by SNIa,
exactly as is Fe in the solar vicinity (see GP2000). This assumption
introduces a time-delay in the production of the bulk of Zn with respect to
 the one of 
other $\alpha$-elements, like S, coming from massive stars. The reason for
that delay is that the bulk of Fe is always produced with some delay in our
models, through SNIa explosions. This explains 
the obtained decline of  $\alpha$/Zn in our models {\it only at
redshifts lower than $\sim$2}. However, physically, this
picture is not correct: it is well known that SNIa do not produce Zn (e.g. 
Iwamoto et al. 1999). The fact that in the solar vicinity Zn behaves
as Fe also at late times is obviously a coincidence. The late Fe production
of SNIa is matched by the Zn production of SNII, which is metallicity dependent
in both the yields of WW1995 as well as in the recent work
of Limongi et al. (2000). 
The yields of WW1995, however, do not reproduce the observed evolution of
Zn/Fe in the Milky Way and  we used Fe as a proxy for Zn; in that way we 
``forced'' a pure metallicity effect (which can, in principle,
manifest itself quite early in zones that evolve rapidly to
high metallicities) to mimic a time-delay effect. In that sense, our treatment
of Zn is not quite correct. Obviously, the nucleosynthesis of Zn should
be carefully scrutinised in future models of massive star nucleosynthesis.

At this point it seems difficult to gain more insight into the nature of DLAs
by simply considering their abundance patterns. We think, however, that
progress can be made through further observations of S in DLAs of high Zn 
column density and of Mn in DLAs of low Zn column density; such observations
would certainly help to establish whether the aforementioned trends are real 
or not.

\section{Summary}

In this work we study the chemical abundance patterns observed in DLAs.
First, we show that there is strong observational evidence for depletion
of the Fe$-$peak elements, obtained when their abundance ratios [X/Zn] are 
plotted as a function of Zn column density. The resulting anticorrelation is
reminiscent of similar depletion patterns obtained in Milky Way clouds as a
function of gas column density. However, the intrinsic gas metallicity 
does not differ much from solar in Milky Way clouds, while it varies by factors
$\sim$ 100 in DLAs. For that reason we argue that 
{\it Zn column density is a more appropriate parameter to measure depletion}, 
since it is proportional (to a first approximation) to the amount of dust along 
the line of sight. Further measurements of [X/Zn] ratios as a function of 
$\mathcal{F}$ = [Zn/H]+log(\N) in DLAs will help to put the anticorrelation 
found here on a firmer basis.

Assuming that DLAs are (proto-)galactic disks, we study their chemical evolution 
with detailed models, making use of metallicity-dependent yields from
massive stars and SNIa; these models reproduce quite successfully a large 
number of properties of spirals at low and high redshift as shown in a series
of papers (Boissier et al. 2000; BP2000, 1999; PB2000a, 2000b). By taking into 
account observational constraints on the gas and metal column densities of DLAs
(namely that log(\N$>$ 20.2 and 18.8$<\mathcal{F}<$ 21) we find that our models
reproduce fairly well (and without  including any depletion
effects) the observed weak evolution of DLA metallicity with
redshift and the corresponding scatter, for several elements detected in DLAs:
S, Si, Mn, Cr, Fe, Ni and Zn.

Encouraged by the success of the model we study  the corresponding predicted
abundance patterns. We find no clear correlation of [X/Zn] with metallicity 
or HI column density. We also find that Si/Zn and S/Zn are supersolar at
high redshifts, while Mn/Zn is subsolar. The latter effect results from 
the metallicity dependent yields adopted here and is also observed in the
Milky Way (see GP2000 and references therein).
The former results from our assumption that [Zn/Fe] = 0 always, assumption 
based on observations in the Milky Way and made because of the poorly
understanding of Zn nucleosynthesis. 

Comparing the observations of [X/Zn] in DLAs to our model calculations
((which do not include the effects of dust depletion) we find evidence for
dust depletion effects in the observations
for Cr, Ni, Fe, Mn and Si, for high $\mathcal{F}>$ 19.8. This depletion
is $\sim$0.4 dex, on average, and comparable to the one observed in
warm halo clouds. However, a difficulty arises with S which is found to be
depleted at high $\mathcal{F}$. If on the other hand it is assumed that the
intrinsic DLA pattern is solar, a problem arises with Mn, which is found to be
depleted at low $\mathcal{F}$. 

We think that currently available data on DLA abundances  do not allow 
to establish the intrinsic abundance patterns of those systems.
Progress can be made through further observations of abundances in DLAs,
in particular those of Mn and S, by taking into account the aforementionned role
of the metal column density. We also note that the conclusions of this
work are largely based on the, presently poorly understood,
nucleosynthesis of Zn; the production of that element should be carefully
studied in new models of massive star nucleosynthesis.

\acknowledgements J.L. Hou acknowledges the warm hospitality of the IAP 
(Paris, France). We are grateful to Patrick Petitjean for useful discussions
and comments on this subject.
This work was made possible thanks to the support of China Scholarship 
Council(CSC) and National Natural Sciences Foundation of China and the NKBRSF19990754.

\def\apj{ApJ}
\def\apjl{ApJL}
\def\apjs{ApJS}
\def\aj{AJ}
\def\aap{A\&A}
\def\araa{ARA\&A}
\def\aapss{A\&AS}
\def\mnras{MNRAS}
\def\nature{Nature}
\def\apss{Ap\&SS}
\def\pasp{PASP}

{}


\begin{thebibliography}{}  

\bibitem[]{} Boiss\'e P., Le Brun V., Bergeron J. \& Deharveng J.M. 1998 \aap, 333, 841
\bibitem[]{} Boissier S., Boselli A., Prantzos N. \& Gavazzi G. 2000, \mnras, accepted
\bibitem[]{} Boissier S. \& Prantzos N. 2000, \mnras 312, 398 (BP2000)
\bibitem[]{} Boissier S. \& Prantzos N. 1999, \mnras 307, 857 (BP1999)
\bibitem[]{} Centurion M., Bonifacio P., Molaro P. \& Vladilo G. 2000, ApJ, 536, 540
\bibitem[]{} Centurion M., Bonifacio P., Molaro P. \& Vladilo G. 1998 \apj, 509, 620
\bibitem[]{} de la Varga A., Reimers D., Tytler D., Barlow T. \& Burles S. 2000 
             astro-ph/0008366
\bibitem[]{} Ferrini F., Molla M. \& Diaz A. I. 1997, \apjl, 487, 29
\bibitem[]{} Gonzalez A., Williams K., Bullock J., Kolatt T.S. \& Primack J.R. 
             2000, \apj, 528, 145
\bibitem[]{} Goswami A. \& Prantzos N. 2000, \aap, 359, 191 (GP2000)
\bibitem[]{} Haehnelt M., Steinmetz M. \& Rausch M., 1998, ApJ, 495, 647
\bibitem[]{} Howk J. C., \& Sembach K., 1999, ApJ, 523, L141
\bibitem[]{} Hou J.L., Prantzos N. \& Boissier S. 2000, \aap, 362, 921
\bibitem[]{} Izotov Y., Scherrer D. and Charbonnel C. 2000, AA, in press (astro-ph/0010643)
\bibitem[]{} Iwamoto K., Brachwitz F., Nomoto K., Kishimoto N., Umeda H., Hix W.R.
             \& Thielemann F. 1999, \apjs, 125, 439
\bibitem[]{} Jimenez R., Bowen D.V. \& Matteucci F. 1999 \apj, 514, L83
\bibitem[]{} Jimenez R., Padoan P., Matteucci F. \& Heavens A., 1998, MNRAS, 299, 123 
\bibitem[]{} Lauroesch J.T., Truran J.W., Welty D.E. \& York D.G. 1996 \pasp, 108, 641
\bibitem[]{} Lindner U., von-Alvensleben U.F. \& Fricke K.J. 1999 \aap, 341, 709
\bibitem[]{} Lu L., Sargent W.L.W., Barlow T.A., Churchill C.W. \& Vogt S.S. 1996 
             \apjs, 107, 475
\bibitem[]{} Matteucci F., Molaro P. \& Vladilo G. 1997, A\&A, 321, 45 
\bibitem[]{} Meusinger H. \& Thon R. 1999 \aap, 351, 841
\bibitem[]{} Mo H.J., Mao S.D.\& White S.D.M. 1998 \mnras, 295, 319
\bibitem[]{} Molaro P., Bonifacio P., Centurion M., D'Odorico S., Vladilo G., Santin P.
             \& Di Marcantonio P. 2000, ApJ, 541, 54
\bibitem[]{} Molaro P., Centurion M.,\& Vladilo G. 1998 \mnras, 293, L37
\bibitem[]{} Pei Y.C., \& Fall S.M. 1995 \apj, 454, 69
\bibitem[]{} Petitjean, P. \& Ledoux, C. 1999 \apss, 265, 483 
\bibitem[]{} Pettini M., Ellison S.L., Steidel C., Shapley A.E. \& Bowen D.V. 
             2000 \apj, 532, 65
\bibitem[]{} Pettini M., Ellison S.L., Steidel C.C. \& Bowen D.V. 1999 \apj, 510, 576
\bibitem[]{} Pettini M., King D.L., Smith L.J. \& Hunstead R.W. 1997a \apj, 478, 536
\bibitem[]{} Pettini M., Smith L.J., King D.L. \& Hunstead R.W. 1997b \apj, 486, 665
\bibitem[]{} Pettini M., Smith L.J., Hunstead R.W. \& King D.L. 1994 \apj, 426, 79
\bibitem[]{} Prantzos N. \& Boissier S. 2000a, \mnras, 315, 82 (PB2000a) 
\bibitem[]{} Prantzos N. \& Boissier S. 2000b, \mnras, 313, 338 (PB2000b) 
\bibitem[]{} Prantzos N. \& Silk J. 1998, \apj, 507, 229 
\bibitem[]{} Prochaska J.X. \& Wolfe A.M. 2000 \apj, 533, L5
\bibitem[]{} Prochaska J.X. \& Wolfe A.M. 1999 \apjs, 121, 369
\bibitem[]{} Savage B.D. \& Sembach K.R. 1996 \araa, 34, 279
\bibitem[]{} Savaglio S., Panagia N. \& Stiavelli M., 1999, in ``Cosmic Evolution
             and Galaxy Formation'', Eds. J. Franco et al., in press ()astro-ph/9912112)
\bibitem[]{} Thielemann F., Nomoto K. \& Yokoi K. 1986, \aap, 158, 17
\bibitem[]{} Timmes F.X., Lauroesch J.T. \& Truran J.W. 1995a \apj, 451, 468
\bibitem[]{} Timmes F.X., Woosley S.E. \& Weaver T.A. 1995b \apjs, 98, 617
\bibitem[]{} Valageas P., Schaeffer R. \& Silk J., 1999, A\&A, 691, 711
\bibitem[]{} Vladilo G., Bonifacio P., Centurion M. \& Molaro P. 2000 \apj, 
             in press (astro-ph/0005555)
\bibitem[]{} Vladilo G. 1999, in ``The evolution of galaxies on cosmological timescales'',
             eds. J.E.Beckman \& T.J. Mahoney, ASP Conf. Ser. vol. 187, 323
\bibitem[]{} Vladilo G. 1998 \apj, 493, 583
\bibitem[]{} Wakker B.P. \& Mathis J.S. 2000 \apjl, accepted (astro-ph/0010045)
\bibitem[]{} Woosley S.E. \& Weaver T.A. 1995 \apjs, 101, 181 (WW1995)

\end{thebibliography}
\end{document}